\documentclass{aastex62}

\shorttitle{oscillatory reconnection}
\shortauthors{Hong et al.}

\begin{document}

\title{Observation of a reversal of breakout reconnection preceding a jet: evidence of oscillatory magnetic reconnection\textbf{?}}

\author{Junchao Hong}
\affiliation{Yunnan Observatories, Chinese Academy of Sciences, Kunming 650216, China}
\affiliation{Center for Astronomical Mega-Science, Chinese Academy of Sciences, Beijing, 100012, China}
\affiliation{Key Laboratory of Solar Activity, National Astronomical Observatories of Chinese Academy of Science, Beijing 100012, China}

 \author{Jiayan Yang} 
 \affiliation{Yunnan Observatories, Chinese Academy of Sciences, Kunming 650216, China}
\affiliation{Center for Astronomical Mega-Science, Chinese Academy of Sciences, Beijing, 100012, China}

 \author{Huadong Chen}
\affiliation{Key Laboratory of Solar Activity, National Astronomical Observatories of Chinese Academy of Science, Beijing 100012, China}
 \author{Yi Bi} 
 \affiliation{Yunnan Observatories, Chinese Academy of Sciences, Kunming 650216, China}
\affiliation{Center for Astronomical Mega-Science, Chinese Academy of Sciences, Beijing, 100012, China}
 \author{Bo Yang}
 \affiliation{Yunnan Observatories, Chinese Academy of Sciences, Kunming 650216, China}
\affiliation{Center for Astronomical Mega-Science, Chinese Academy of Sciences, Beijing, 100012, China}
 \author{Hechao Chen}
\affiliation{Yunnan Observatories, Chinese Academy of Sciences, Kunming 650216, China}
\affiliation{Center for Astronomical Mega-Science, Chinese Academy of Sciences, Beijing, 100012, China}
\affiliation{Graduate School of Chinese Academy of Sciences, Beijing, P.R. China}

\correspondingauthor{Junchao Hong}
\email{hjcsolar@ynao.ac.cn}

\begin{abstract}
 
Recent studies have revealed that solar jets involving minifilament eruptions may be initiated under the well-known magnetic-breakout mechanism. Before or just at the onset of those jets, there should be a current sheet, where breakout magnetic reconnection takes place, between open fields and the outside of the jet-base arcade carrying minifilament in its core.  In this paper we present a jet produced by eruption of two minifilaments lying at the jet base.  A current sheet is directly detected near the jet base before the onset of the eruption, suggesting the magnetic-breakout mechanism. However, we further find that the current sheet undergoes a transition. The current sheet first shortens to zero in length, but then lengthens towards an orthogonal direction relative to its initial orientation. The change of the current sheet gives rise to a reversal of the breakout reconnection, as the inflow and outflow regions before the transition become the outflow and inflow regions after the transition, respectively. We therefore propose that this observation provides evidence for the so-called oscillatory reconnection which is defined by a series of reconnection reversals but not yet proved to exist in real plasma environment of the solar atmosphere. 
\end{abstract}

\section{introduction}
 Solar jets are transient, collimated plasma ejections that escape away from the low solar atmosphere along open or far-reaching magnetic fields. They have been observed to be ubiquitous across the Sun  surface and studied extensively over the last several decades (see the recent reviews \citealt{2016Innes, 2016Raouafi}). Theoretically, it is widely accepted that jets are produced by magnetic reconnection at a magnetic-null region between a closed field and surrounding open field \citep[e.g.][] {Shibata1992,Yokoyama1995,2005Archontis,2007Jiang,2008Chenhd,2015Pariat,2016Pariat,2017Ni}. The magnetic null is susceptible to collapse to form a current sheet where bursty reconnection can takes place \citep{Antiochos1990,Priest2000}. Observationally, the magnetic configuration with a magnetic-null point prior to a jet has been inferred from the specific fan-spine shapes of jet emissions in the corona, the circular flare ribbons in the chromosphere at jet base, and the so-called embedded bipolar fields in the photosphere \citep{2012Wanghm,2012Zhangqm,2017Lihd,2017Mccauley}. However,  the current sheet between the closed field and open field  has rarely been observed directly at the onset of a jet.
 
 Recent observations have revealed that solar jets are accompanied by minifilament \citep{Wangjx2000} eruptions in the manner like large coronal mass ejections (CMEs) by large-scale filament eruptions \citep{2015Lix,Sterling2015,Sterling2016,Hong2016,Hong2017,2016Panesar,2017Panesar,2018Panesar,2017Zhangyz,2012Shena,2017Shen,2018Moore}. \cite{Sterling2015} select 20 polar jets randomly which are recorded  in coronal images of X-ray and extreme ultraviolet (EUV) wavelengths simultaneously.   They find that for each example an EUV minifilament erupted upward from the pre-jet base, producing a X-ray jet above the base and leaving a flare-like brightening near the base.  Using the joint observations from  New Vacuum Solar Telescope (NVST; \citealt{Liuz2014}) and {\it Solar Dynamic Observatory} ({\it SDO}; \cite{Pesnell2011}) , \cite{Hong2016} report in detail how a standard jet seen in EUV wavelength  is initiated by a minifilament eruption detected at H$\alpha$ emission line.  The erupting minifilament forced the external closed field enveloping itself  to reconnect with the long coronal loops nearby. As a result, an jet along the loops was produced by that external reconnection.  The external reconnection for the jet production corresponds to the well-known breakout reconnection for CMEs involving large filament eruptions \citep{1998Antiochos,1999Antiochos,2012Shenb,2016Chen}.
 Very recently, \cite{2018Moore}  find that 6 of 15 polar X-ray jets they studied are plausibly initiated due to the breakout reconnection in each minifilament eruption . If the data has enough resolution and  the viewpoint is proper,  a current sheet, the site for breakout reconnection, should be seen between  the jet-base explosive close field and the open field for those jets. 

Via three-dimensional (3D)  magnetohydynamic simulations, \cite{2018Wyper} present a breakout model for the minifilament-jet events, as a small-scale extension of the well-known breakout mechanism for large CMEs.  The basic magnetic setup  consists of the background open fields and an embedded bipolar field, separated by a null point.  They give a persistent shearing motion at the bipolar footpoints in their simulations. Over time, a sheared filament channel forms and expands in the core of the bipolar filed, and  the magnetic pressure within the bipolar field increases constantly. Sequently, the breakout current sheet is built  because the null point above the bipolar field becomes increasingly compressed.   As the breakout reconnection open the strapping fields above the minifilament to some extent, the minifilament  erupts upward inevitably towards the breakout current sheet. The field (possibly a flux rope) holding and threading the rising minifilament continues to reconnect with the open fields, producing an apparent jet spire. Meanwhile, the flare reconnection under the erupting minifilament occurs impulsively, accelerating the escape of the ejecta and making miniature flare-like brightening at the jet base.  \cite{2018Wyper} suggest that the breakout mechanism for jet generation is robust in despite of varying inclinations of the background open fields.  In addition, the breakout behavior is not sensitive to  whether the minifilament in the core of the bipolar field is held by a sheared arcade or true flux rope.
 
In this paper we present a jet produced by successive eruptions of two minifilaments lying at the pre-jet base. The jet displays typical fan-spine topology where the breakout reconnection occurs plausibly at the null point, evidenced by a current sheet directly detected before the onset of the jet. However, the current sheet takes an interesting behavior:  It shortens its length to zero along its orientation, and then lengthens  towards an orthogonal direction relative to its initial orientation. The entire process is plausibly in self generation. The change of the current sheet indicates that the inflow and outflow regions of the breakout reconnection interchange each other.  This gives one a reminiscent of the ``oscillatory reconnection"  featured by a series of reconnection reversals in a self-consistent manner, which is first reported in numerical simulations  \citep{2009Murray} but not yet found in observations.

\section{observations and data}
The jet is recorded by {\it SDO} and the Interface Region Imaging Spectrograph ({\it IRIS}; \citealt{2014Depontieu}) simultaneously.  The jet started at about 15:26 UT, peaked at around 15:34 UT, and became faint after 15:50 UT on 2014 November 16 near the Active Region AR12209 (S15E32) on solar disk. 

{\it SDO} provides full-disk imaging data by the onboard telescopes Atmospheric Imaging Assembly (AIA; \citealt{2012Lemen}) and Helioseismic and Magnetic Imager (HMI; \citealt{2012Schou}). AIA images the Sun at UV-EUV wavelengths with cadences of 24-12 s and pixel sizes of 0.$\arcsec$6. HMI gives the photospheric magnetograms with 45 s cadence and 0.$\arcsec$5 pixel size. In the jet, the AIA  images at EUV wavelengths (e.g. 131 \AA\ )  image plasmas at the coronal temperature (above 10$^5$ k),   and thus are used to reveal the fan-spine shape of the jet.  The current sheet associated with the jet  can be observed at all of the EUV wavelengths of AIA including the He II emission line at 304 \AA\ , and six Fe emission lines at 171,193,211,335,131,94 \AA\ . Therefore, plasmas within the current sheet can be diagnosed with the almost simultaneous observations of the six AIA Fe lines via  differential emission measure (DEM) analysis \citep[e.g.][]{2011Aschwanden,2012Chengx,2018Su}. The AIA UV channel 1600 \AA\ are dominated by chromospheric continuum and transition-region line emission, and thus are used to show the circular flare ribbon and remote brightening in the fan-spine jet, and time profile of the flare emission. HMI line-of-sight magnetograms are used to inspect the magnetic field configuration and evolution in the jet source region.

{\it IRIS}  imaged the jet source region from 14:54 UT to 15:34 UT by its slit-jaw camera at 1330, 2796, and 1400 \AA\ . The observational period covers the pre-jet evolution and the growing phase of the jet.  The 1400 \AA\ slit-jaw images (SJIs) are used here with a cadence of 37 s and a high spatial resolution of 0.$\arcsec$33. The SJI 1400 \AA\ images reflect plasma of bright chromospheric continuum and Si IV transition-region line emission, and thus are easily aligned to the AIA 1600 \AA\ images. Due to the extreme high resolution of {\it IRIS}, the current sheet is very discernibly detected to undergo the changes of its length and orientation. In addition, the  two minifilament at the pre-jet base are also found by  {\it IRIS}/SJI observation.

\section{results}
Figure 1 presents the overview of the jet with the {\it SDO} data. A time sequence of the interesting region for the jet is displayed in panels a1-a4 with AIA 131 \AA\ images at four different moments . At 15:18 UT and 15:22 UT, the jet spire was not seen and the jet-base region (outlined by the orange boxes) did not become bright. At 15:29 UT, the jet base became a compact, bright dome and a long, bright jet spire stemmed from the top of the bright dome. The jet spire was arose along  preexisting, far-reaching coronal loops and a bright brightening appeared at the far ends of those loops. The AIA 1600 \AA\ image in panel d further detected a circular bright ribbon in the jet base besides the remote brightening. Combining with the magnetogram (panel e), it is know that the circular ribbon is located at some scattered positive polarities that surround a negative polarity (N) at center. There are two main positive patches, labeled as P1  on the north of N  and P2  on the south of N. These features point to an existence of magnetic configuration of the fan-spine topology \citep{2009Masson,2014Masson}.  At 15:38 UT, the jet spire developed multi separate strands and broaden to about the width of the jet base, displaying the typical morphology of blowout jet \citep{Moore2010}.  

Two linear-shaped bright structures were seen near the jet base before the onset of the jet. They are the line segment C1 at 15:18 UT and  C2 at 15:22 UT, as shown by the close-up views of the jet base in panels g and h respectively. Each end of C1/C2 presents a bifurcate structure, connecting a very bent curve of bright structure. This pattern gives a reminiscent of the classical 2D picture of magnetic reconnection, indicating that C1/C2 is the projection of a current sheet \citep{1957Parker,1958Sweet,1964Petschek}. The current sheet separates two sets of anti-parallel fieldlines at its two sides as the inflow regions before reconnection, and connects  the reconnected fieldlines at its two ends as the outflow region after reconnection.  Recently, small current sheets between  two sets of anti-parallel loops are reported in several papers \citep{2015Yangsh, 2016Yangsh,2016Xue,2018Xue,2018Yangb}.
For the present jet, the existence of C1/C2 indicates the happening of the so-called interchange reconnection between closed field and ``open" field \citep{2002Crooker}.

According to the linkage of C1/C2 with the surrounding bent bright structures and the overlaid magnetic fields in panels g/h, we can infer the connectivity of the inflow and outflow fieldlines for the interchange reconnection at C1/C2.  For C1, magnetic reconnection should take place between the closed fieldline connecting P1 to N and the open line along the jet spire rooted at P2. As a result, one outflow fieldline is open with footpoint at P1, as traced out by the bent bright structure contacting the left end of C1; and the other outflow fieldline is a close loop connecting P2 to N. Again, for C2, magnetic reconnection should occur between the closed fieldline connecting P2 to N and the open line along the jet spire rooted at P1. Consequently, one outflow fieldline is open with footpoint at P2 and the other one is the closed loop connecting P1 to N. Comparing C1 with C2, it is concluded that (1) the orientation of C1 is orthogonal to that of C2, as C1 is roughly horizontal but C2 is roughly vertical; (2) the inflow region and the outflow region of C1 is, respectively, the outflow region and the inflow region of C2. By overplotting C1/C2 in panel a3, it is seen that C1/C2 are located at the joint-point of the jet spire and the fan-shaped dome. Thus C1/C2 seem to be built in the vicinity of the null point in the fan-spine magnetic configuration. The bright structures that connect the ends of C1/C2 to N and open possibly trace out the inner and outer spines, while those linking the ends of C1/C2 with P1 and P2 indicate the projection of separatrix of the fan dome.

Figure 2 presents the results of DEM analysis on C1 at 15:18 UT and C2 at 15:22 UT.  The current sheets are simultaneously imaged by multi wavelengths of AIA, as shown in the three-passband composite images for 171,211, and 131 \AA\ (C1 in panel a1 and C2 in panel a2). Via DEM method, the temperature maps  for C1/C2 are constructed as displayed in panels b1/b2, and the emission measure maps in panels c3/c4. Apparently, it is seen that the current sheets are hotter and denser than ambient regions. 

The DEM distributions of plasmas inside C1/C2 are further derived as shown in Figure 2(e/f). The results are computed by using the mean digital numbers (DNs) over an area (outlined by the boxes in panels a1/a2) subtracted by a background value determined from the nearby quiet regions. With 100 Monte Carlo simulations of the DEM inversion calculation, it is seen that the DEM curves are well constrained in a broad temperature range  although poorly constrained in a few temperature bins. C1 and C2 have almost similar DEM distribution and are multi-thermal as signed by the broad distribution from $log T=5.5$ to $log T=7.5$. Both them have the peak temperature at $log T=6.2$ ($\sim$1.6 MK).
The total emission measure (EM) and DEM-weighted mean temperature of C1/C2 are further calculated according to the expresses ${\mbox{EM}}={\int}{\mbox{DEM}}(T)dT$ and $T={\int}{\mbox{DEM}}(T)TdT/{\mbox{EM}}$.  C1 has an averaged temperature of $\sim$7 MK, slightly hotter than that of 6.8 MK for C2.  But C2 with an EM of  2.5x10$^2$$^8$ cm$^-$$^5$, is denser \textbf {(or has a greater extent along the line of sight)} than C1 with an EM of 1.4x10$^2$$^8$ cm$^-$$^5$. 

The jet source region is simultaneously watched by {\it IRIS}. With the unprecedented high resolution of {\it IRIS}/SJI 1400 \AA\ images, it allows one to discuss the evolution of C1/C2 on detail. Figure 3 shows    a successive process from C1 to C2. At 14:55 UT, no current sheet was seen at the pre-jet base, but two minifilaments, F1 and F2, were found lying at the base. With the overlaid magnetic field, it is shown that F1 resided along the neutral line between P1 and N while F2 along the neutral line between N and P2.  This also indicates the existence of an X-type null point above the two minifilament, similar to the situation that an X-type null point above the  twin magnetic arcades under a coronal pseudo streamer \citep{2007Wangym,2013Bi,2015Yangjy}.   The yellow arrow points to the approximate location of the expected X-type null point where the current sheet C1 appeared later. 

C1 was faintly seen around 14:58 UT at the first time in 1400 \AA\ images. It was still faint at 15:09:26 UT but later became discernible gradually. When C1 got most long at 15:17:32 UT,  jet-like flow emitting from the left end of C1 was also discernible.  At this time, C1 was estimated to have an apparent length of $\sim$2$\arcsec$.3  and a width of $\sim$0$\arcsec$.6. Four domains, registered with ``A", ``B", ``D", and ``E" in the 15:17:32 UT image, were separated by C1. According to the classic model of magnetic reconnection, it is known that the domains A and B at each side of C1 represent the inflow regions for the reconnection while D and E indicate the outflow regions at the two ends of the current sheet. From the 15:17:32 UT image with overlaid magnetic contours, it is  inferred that the inflow region A is occupied by open field rooted near P2 while the inflow region B contains closed field linking P1 to N. Therefore, interchange reconnection took place between the open field in A and the closed field in B \citep{2002Crooker}. Consequently,  new open field along which the jetlike flow propagated outward, was formed to be rooted at P1 in the outflow region D. New closed field linking N to P2 was piled in the outflow region E. 

Note that the reconnection at C1 actually removed the  magnetic arcade (P1-N) that covers the underneath minifilament F1.  In this respect, the interchange reconnection corresponds to the so-called breakout reconnection as a trigger for many large filament eruptions \citep{1998Antiochos,1999Antiochos}. However, the breakout reconnection did not trigger F1 to erupt immediately. Instead, the breakout current sheet, C1, started to shorten itself  at one moment after 15:17:32, as its outflow regions D-E was approaching each other. It is seen that C1 at 15:20:01 UT was shorter than it at 15:17:32 UT. To 15:21:16 UT, the outflow regions D-E  met each other, and an X-type structure instead of C1 was built separating the four regions A, B, D, and E.  In the next one minutes,   the outflow regions D and E have further compressed each other, and the inflow regions A and B have been pull away from each other. As a result, a new current sheet, C2, vertical with C1, appeared to separate D and E as its inflow region and A and B as its outflow region (see the 15:22:31 UT image).  C2 has been further compressed to be thinner (as seen at 15:25:38 UT), ultimately reaching an length of $\sim$2$\arcsec$.5 and a width $\sim$0$\arcsec$.7.  Similar to that C1 did as the breakout current sheet above F1, C2 played a role of the breakout current sheet above F2. 
The transition from C1 to C2 demonstrated a reversal of the breakout reconnection. This further indicates that the breakout reconnection took place in the manner of oscillatory magnetic reconnection  \citep{2009Murray, 2010Archontis,2012Mclaughlin, 2017Thurgood}. A detailed discussion is present in the last section.

A time-slice plot, as shown in the last frame of Fig. 3, was constructed by the slit along the long arrow in the 15:22:31 UT image of Fig. 2. It is to display the duration of C1 and C2, since the slit passes across C1 and along C2.  C1 lasted about 23 minutes from 14:58 UT to 15:21 UT, while C2 lasted around 7 minutes from 15:22 UT to 15:29 UT. The transition from C1 to C2 took place in one minute between 15:21 UT and 15:22 UT.

Although F1 did not erupt immediately because of the breakout reconnection at C1, F2 started to brighten and rise  as soon as the appearance of the breakout current sheet of C2.  Fig. 4 presets the eruptive phase of the jet.  At 15:27:30 UT, F2 was already rising up as a bright arch-like feature seen in  the 1400 \AA\ snapshot of Fig.4 (a1), but F1 was still stable as a dark feature. The exact time of the start of the rise  could not be determined due to F2's  very small size and projection mask. However, F2 seems to be activated to be bright immediately after the appearance of C2,  suggested by the 15:22 UT 131 \AA\ image of Fig.4(b1). From panel a1 to a2 , it is seen that F2 expanded itself and approached to C2, keeping its arch shape. Meanwhile,  brightening appeared  between the magnetic polarities N and P2, similar to the initial flare brightening underlying a large-scale erupting filament. Circular brightening also appeared along the ambient positive polarities close to P2 , indicating the enhance of the null-point/breakout reconnection at C2.  A narrow, hot jet spire was thus been aroused  at the same time, as shown by the 131 \AA\ image in pane b2. 

The explosive change took place when F2 reached C2, after which F2 was not seen but a 1400 \AA\  jet spire was launched, leaving remarkable brightening in the base (panel a3).   Base on the intertwined substructure of the jet around 15:29 UT, it is conjectured that F2 should be broken by the null-point/breakout reconnection. As a result, its twist and  cool material was transferred into the jet body \citep{2018Wyper}. We further found that F1  was also ejected out approximately after 15:30 UT,  although the start of its eruption was obscured by the eruption of F2.   Due to the joint action of both minifilament eruptions, the jet has grown a wide spire, accompanied by the development of a circular bright ribbon  at the jet base  (see panels a2 to a5). Examples for a filament eruption triggered by the adjacent eruption of another filament are also reported by other authors \citep[e.g][]{2011Torok,2014Sterling}.  \cite{2018Liting} also discussed the detailed development of circular flare by successive eruptions of two minifilament in another event. 

After 15:34:27 UT, the jet has been only observed by {\it SDO} while {\it IRIS} ended its observation on this region. The jet/minifilament eruptions continued to propagate plasmas along its path as shown by  the 131 \AA\ images of Fig 4(b3-b4). It decayed gradually after 15:40 UT and was almost invisible after 16:00 UT. However, following the later phase of its evolution, we found that a bright linear feature, pointed by the yellow arrows, seems to connect the jet spire to the top of the miniature flaring arcade at the base. This linear feature is likely a flare current sheet in the wake of the eruption of F1, analog to the large current sheet in the wake of CMEs \citep[e.g.][]{2015Lin,2016Li,2017Mei,2018Yan}. Therefore, we observed both type of current sheets in a single jet event, including the breakout current sheet before the onset of the jet and the the flare current sheet in the wake of jet. This result further suggests the self-similarity among the narrow, small jets and the wide, large CMEs that is triggered under the breakout mechanism \citep{2018Wyper}. 

Another evident characteristic of the jet is the photospheric magnetic cancellation,  which is observed in the source region from before to after the jet by HMI. This took place at the neutral line between negative flux N and the surrounding positive flux (mainly P1 and P2), in which F1 and F2 also erupted.  Fig. 5a shows the time sequence of HMI magnetograms for the jet source region. It is  seen that the areas of magnetic polarities P1, N, and P2 have decreased dramatically from 14:53 UT to 15:49 UT due to the cancellation among them.  During the cancellation, the reversal of the breakout reconnection was built above the source region, as sketched  by the  superimposed  axis of F1 /F2, the current sheets C1/C2 and their connectivity to P1, P2, and N.   Fig.5b shows the time profile of magnetic flux of N and the 1600 \AA\ light curve of the source region. It is seen that the magnetic flux continuously decreased through the time of the breakout phase when C1/C2 were present and the jet phase when F1/F2 were erupting. Two peaks in the 1600 \AA\ light curve are found during the jet phase, which are possibly resulted from the successive eruption of both minifilaments \citep[e.g][]{2018Wangy}.  The relationship among the reversal of the observed breakout reconnection, the minifilament eruptions, and the magnetic cancellation will be further discussed in next section.

\section{ discussion and summary}

Solar jets are often found to be made by the explosively eruption of pre-jet base arcade. In general, this eruption is resulted from an eruptive structure seen as a minifilament \citep[e.g.][]{Hong2016} or  a micro-sigmoid \citep{2010Raouafi} or even a small flux rope \citep{2017Zhu,2018Joshi}. The production of solar jets is thought to be produced by the interaction of these eruptive structures with open fields or far-reaching loops \citep{Sterling2015,Hong2016,2018Moore,2018Wyper}. Here, we report a jet that originates from successive eruptions of two minifilaments in the pre-jet base.  One minifilament, F1, lies along the neutral line between positive magnetic flux P1 and negative flux N, while the other one, F2, along the neutral line between N and P2. Magnetic field at the pre-jet base should mainly consist of twin magnetic arcades (respectively cover F1 and F2) tied with a X-type null point above \citep{2007Wangym}.  The breakout reconnection/current sheet is definitely found  before the eruption, which is plausibly built at the X-type null point.  The jet is triggered when F2 first erupts toward the breakout current sheet and participates in the breakout reconnection. F1 erupts immediately following the breaking of F2. Similar to large filament eruption,  a small flare current sheet appears in the wake of  the eruption of F1, which connects the jet spire with the miniature flare arcade in the jet base. Both F1 and F2 are broken and ejected into the jet body. Therefore, the present jet suggests the breakout model for solar jets proposed by \cite{2018Wyper}. This result is also consistent with the observational inspection about the trigger for polar X-ray jets by \cite{2018Moore}.

Magnetic cancellation is also observed at the jet base, which mainly occurred at the neutral lines under F1 and F2. It is well known that magnetic cancellation plays an important role in triggering different scale of filament/jet eruptions \citep[e.g][]{Hong2011,Hong2014,2016Panesar,1992Moore,2012Huang}.  The observed jet should be resulted from the joint action of both of the breakout reconnection above the base arcade  and the magnetic cancellation under the minifilaments. Just before the onset of the jet, F1 does not erupt immediately following the breakout reconnection at C1 while F2 gets bright and rises up as soon as the appearance of C2. We conjecture that this is possibly because the magnetic cancellation under F1 did not disturb F1 as enough as that disturbs F2. This point needs to be verified in future work.

 Properties of this jet, including the breakout reconnection, the minifilament eruptions, and the magnetic cancellation, are similar to those commonly observed in large scale CMEs. Thus, in line with other literatures \citep[e.g.][]{2010Schrijver,2010Raouafi,2017Wyper}, the observations suggest again that multi-scale (from jets to CMEs) of solar eruptive activities are self-similar in terms of observational properties and triggering mechanisms.  
 
 However, the breakout reconnection undergoes a reversal, indicated by the transition of the breakout current sheet from C1 to C2.  C1 is orthogonal to C2. The current sheet displays as C1  about 23 minutes.  Then it shortens in length to zero and grows out again as C2. The transition from C1 to C2 takes about 1 minute. After that, C2 lasts about 7 minutes until the start of the jet. As a result, the inflow and outflow regions around C1 become the outflow and inflow regions around C2, respectively.  The breakout reconnection  thus occurs a reversal, leading to the effect that weakens the magnetic constraint on F1 become that weakens the magnetic constraint on F2. This helps F2 to lift off away from balance. 

The reversal of magnetic reconnection is first found in solar atmosphere. The occurrence of reconnection reversal is possibly a self-consistent behavior \citep{2009Murray}. Using a 2.5-dimensional numerical simulation, \cite{2009Murray} first report ``oscillatory reconnection" that is characterized by a series of reconnection reversals initiated in a self-consistent manner.  The reconnection is first built between an emerged bipolar arcade and ambient open fields in their simulation. Then the reconnection lasts in distinct bursts. The inflow and outflow regions of one burst of reconnection become the outflow and inflow regions in the following burst of reconnection. The magnetic system finally settles toward equilibrium through consecutive bursts of reconnection.  They argues that oscillatory reconnection occur if the outflow regions are quasi-bounded during each burst of reconnection. The reconnection reversal occurs because the gas pressure in the bounded outflow regions increases above the level of that in the inflow regions. For the present jet, the outflow regions around the current sheet C1 are actually quasi-bounded,  thus the reversal occurs and the current sheet changes from C1 to C2. In turn, the observed reversal of reconnection provides the direct evidence  of oscillatory reconnection proposed by \cite{2009Murray}. However, different from the simulation, we do not observe two or more reversals of reconnection that keep the magnetic system stable. Instead, the magnetic fields in the jet base become unstable and erupt out only after one reversal.  This is because of the release of free energy stored in the magnetic fields of the minifilaments in the jet base.  \citet{2014Zhangqm} reports repeating magnetic reconnection in a coronal bright point, which is likely a case of the oscillatory reconnection. More observations are needed to understand deeply the nature of the reversal of magnetic reconnection and confirm if the reconnection reversal is popular in solar atmosphere.

In summary, the present observations directly image the evolution of the breakout current sheet that precedes the eruption of a minifilament-jet event. The breakout current sheet has an apparent length less than 3$\arcsec$ and width less than 1$\arcsec$.   Particularly, according to the transition of the current sheet from  the horizontal C1 to  the vertical C2, we conclude that the breakout reconnection undergoes a reversal in the vicinity of a  potential coronal null point.  The entire evolution of the current sheet can  be detected in images of multi-wavelength emissions including AIA EUV and IRIS transition-region lines. Via DEM analysis, C1/C2 are found to be multi-thermal in nature, with the mean temperatures of about 7/6.8 Mk, the peak temperature at 1.6 Mk, and the EMs  of 1.4/2.5 x 10$^2$$^8$ cm$^-$$^5$. Similar process referring to such reconnection reversal are only be found in numerical simulations for oscillatory reconnection as before \citep{2009Murray, 2010Archontis,2012Mclaughlin, 2017Thurgood}. 

\acknowledgments
 The data used here are courtesy of the {\it IRIS} and {\it SDO} science teams. 
This work is supported by the Natural Science Foundation of China under grants 11503081, 11633008, 11703084, 11333007, 11573012, and 11503082,  and by the CAS programs ``Light of West China'' and ``QYZDJ-SSW-SLH012'', and by the grant associated with the Project of the Group for Innovation of Yunnan Province.

\clearpage
\begin{figure}
\plotone{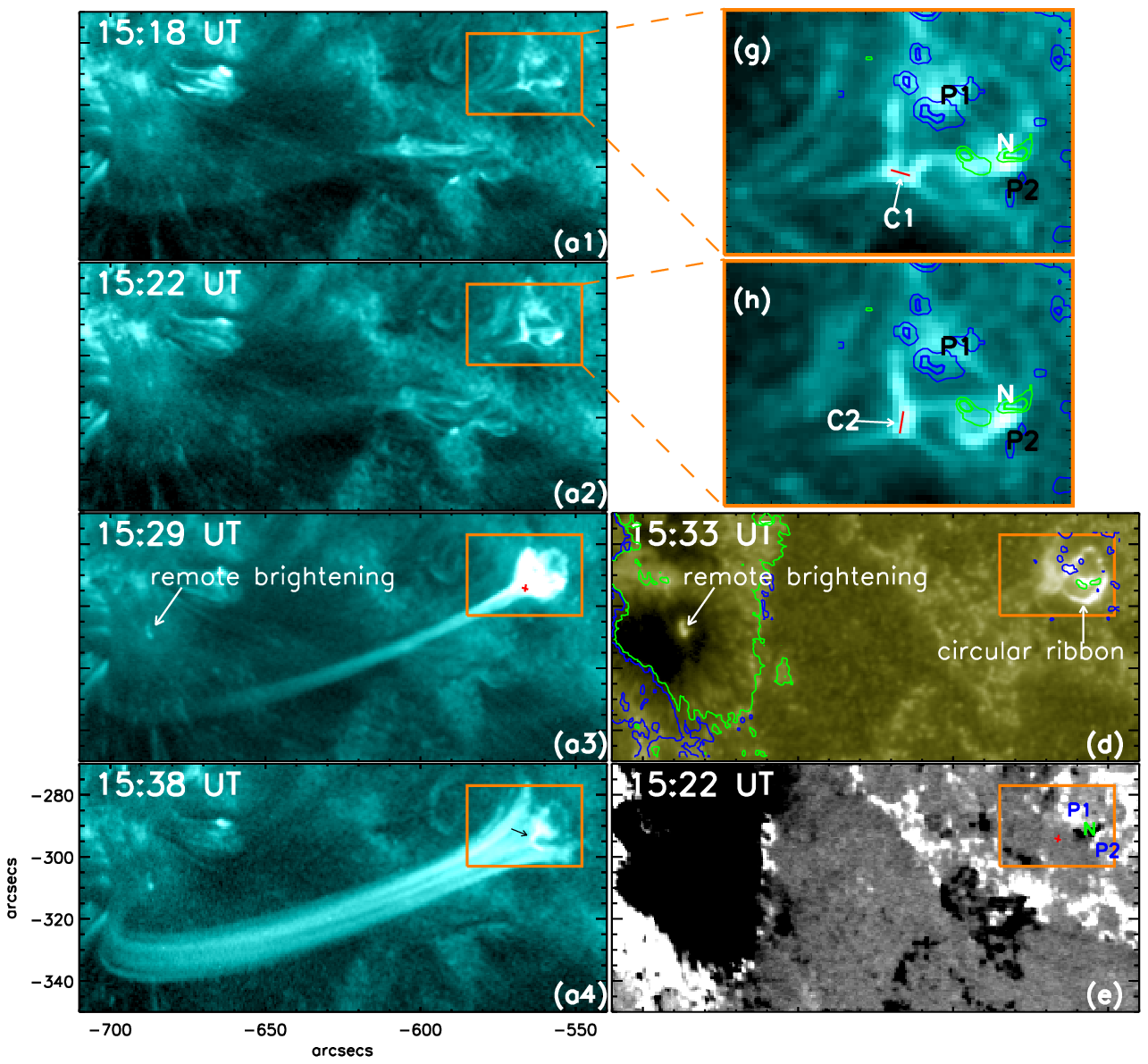}
\epsscale{1.}
\caption{The present jet is fully detected by {\it SDO}. Panels a1-a4: AIA 131 \AA\ images showing the initiation and growth of the jet. Panels g and h: the close-up views within the orange boxes in panels a1 and a2, respectively. To enhance the visibility of the possible current sheets C1/C2, the present 131 \AA\ images are the average of five consecutive snapshots in one minute of the 12s-cadence AIA 131 \AA\ snapshots. Panel d:  AIA 1600 \AA\ image showing the circular ribbon and remote brightening during the jet. Panel e: HMI line-of-sight magnetogram showing the photospheric magnetic fields associated with the jet. The magnetic field in the source region of the jet consists of a negative patch (N) roughly surrounded by some scattered positive patches (mainly P1 and P2).  Green/blue contours in panels g and h are the levels of the HMI magnetic field at (-200, -80)/(200, 80) G, while in panel d at -100/100 G. }
\end{figure}

\begin{figure}
\epsscale{1.}
\plotone{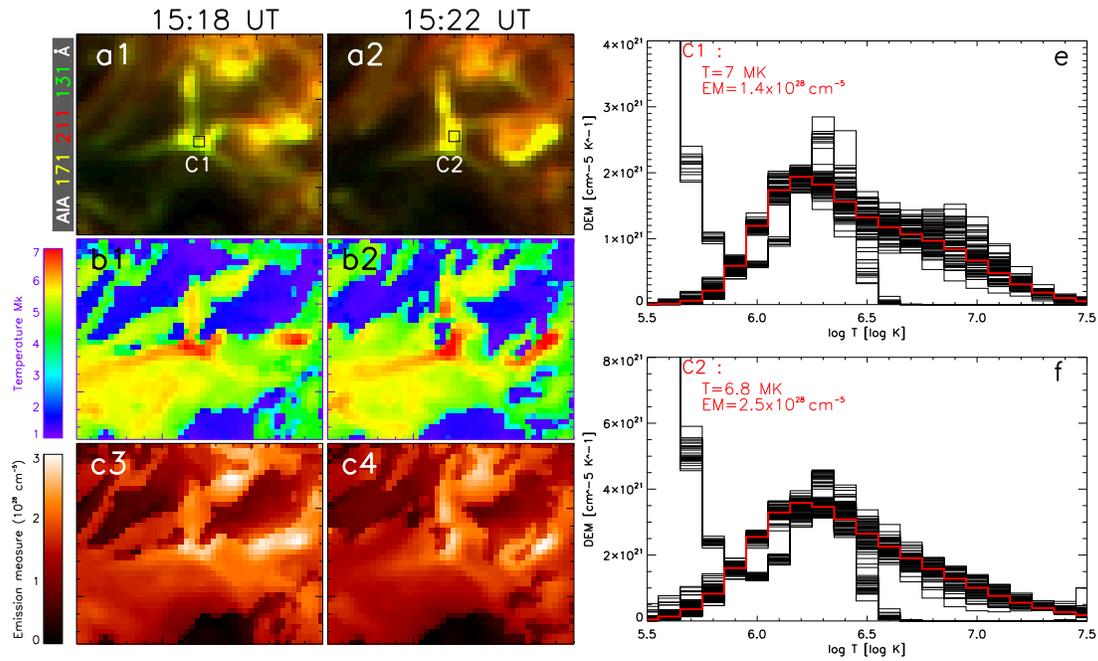}
\caption{The sheet structure of C1/C2 shown in three-passband composite image for AIA 171, 211, and 131 \AA\  (panels a1/a2 ), and in temperature (panels b1/b2) and emission measure (panels c1/c2) maps constructed via DEM analysis. Panels e and f give the averaged DEM distributions of two regions outlined by the small boxes with sizes of 1$\arcsec$.2 at the sheets C1/C2, respectively. The red curves represent the best-fit DEM curves while the piled black curves around the red curve in each panel are 100 Monte Carlo simulations as an estimation of the DEM inversion errors.}
\end{figure}

\begin{figure}
\epsscale{1.}
\plotone{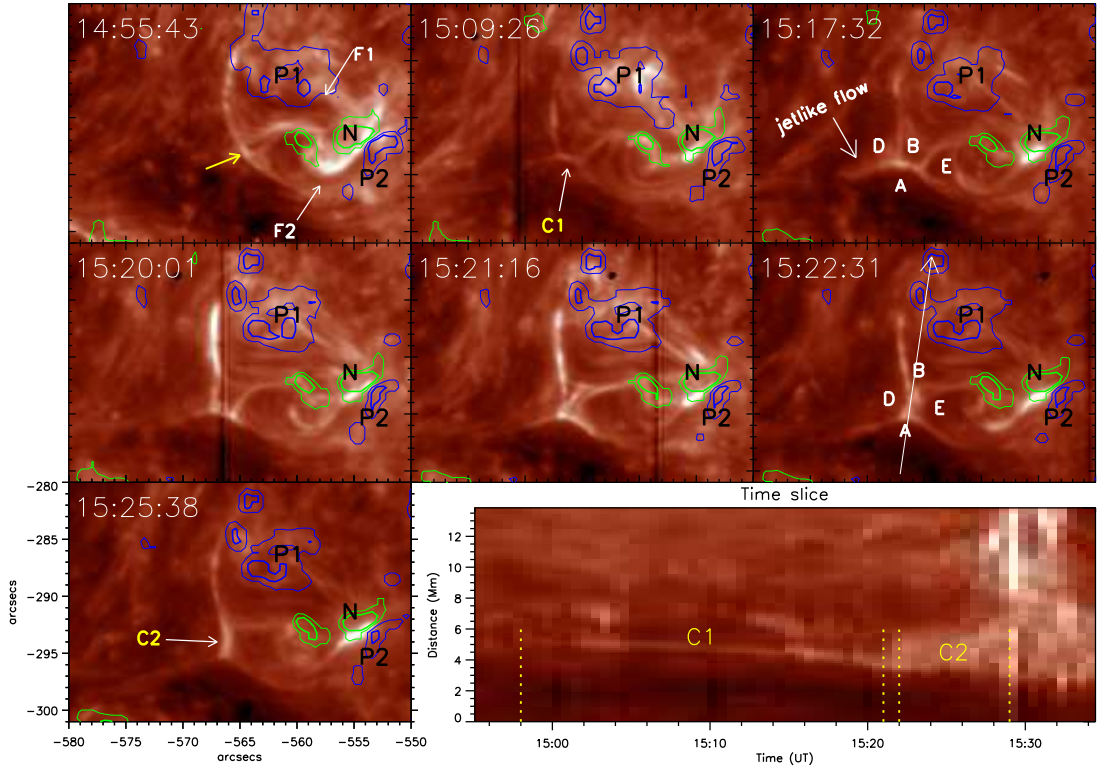}
\caption{The detailed evolution of the current sheets C1/C2 before the jet in the {\it IRIS}/SJI 1400 \AA\ images.  In the first panel, F1 represents the pre-jet minifilament lying along the magnetic polarity inversion zone between P1 and N, while F2 represents the other one between P2 and N. The letters A, B, D, and E indicate four domains around the recconntion site, i.e. C1/C2. In the right bottom panel is a 1400 \AA\ time-slice plot which is made from slits along the long arrow. The yellow dotted lines on the time-slice image indicate the duration of C1 and C2.  The blue (green) contours indicate magnetic field strength at levels of 80, 200 (-80, -200) G. }
\end{figure}

\begin{figure}
\plotone{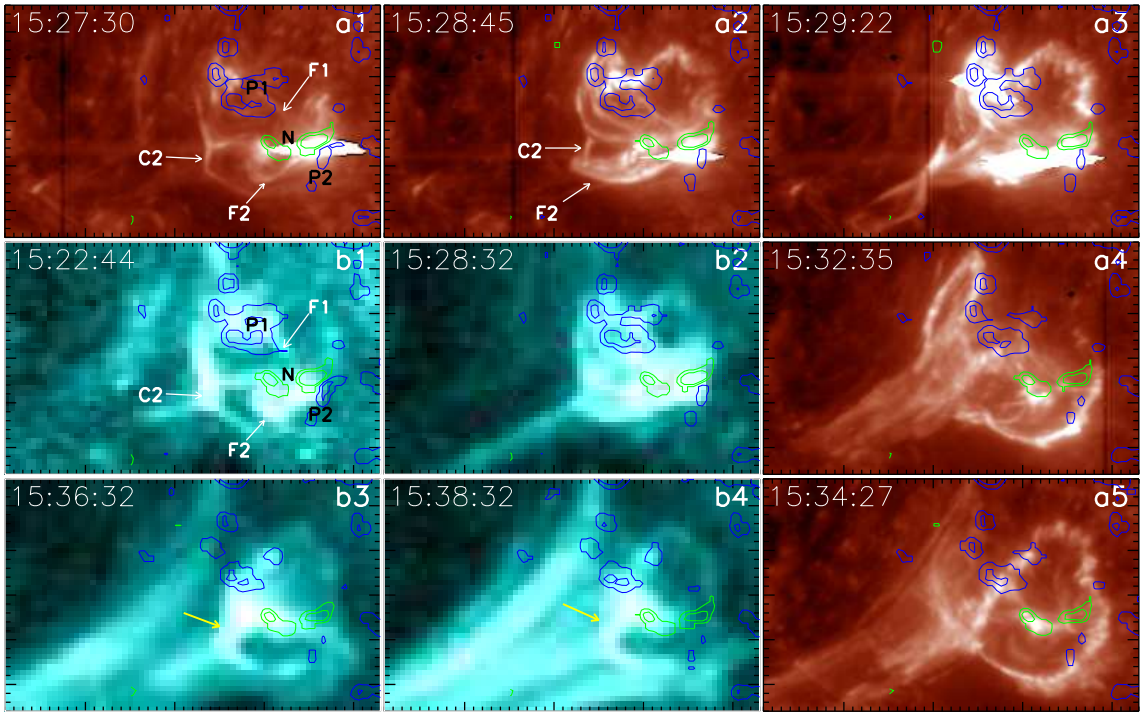}
\caption{Panels a1-a5: The onset and progression of the minifilament-jet eruption in  {\it IRIS}/SJI 1400 \AA\ images.  Panels b1-a4:  The jet in the 131 \AA\  images with the same FOV as that of the 1400 \AA\ images. The green/blue contours indicate field strength at levels of 80, 200 (-80, -200) G. The yellow arrows point to a possible flare current sheet in the wake of the jet.}
\end{figure}

\begin{figure}
\epsscale{1.2}
\plotone{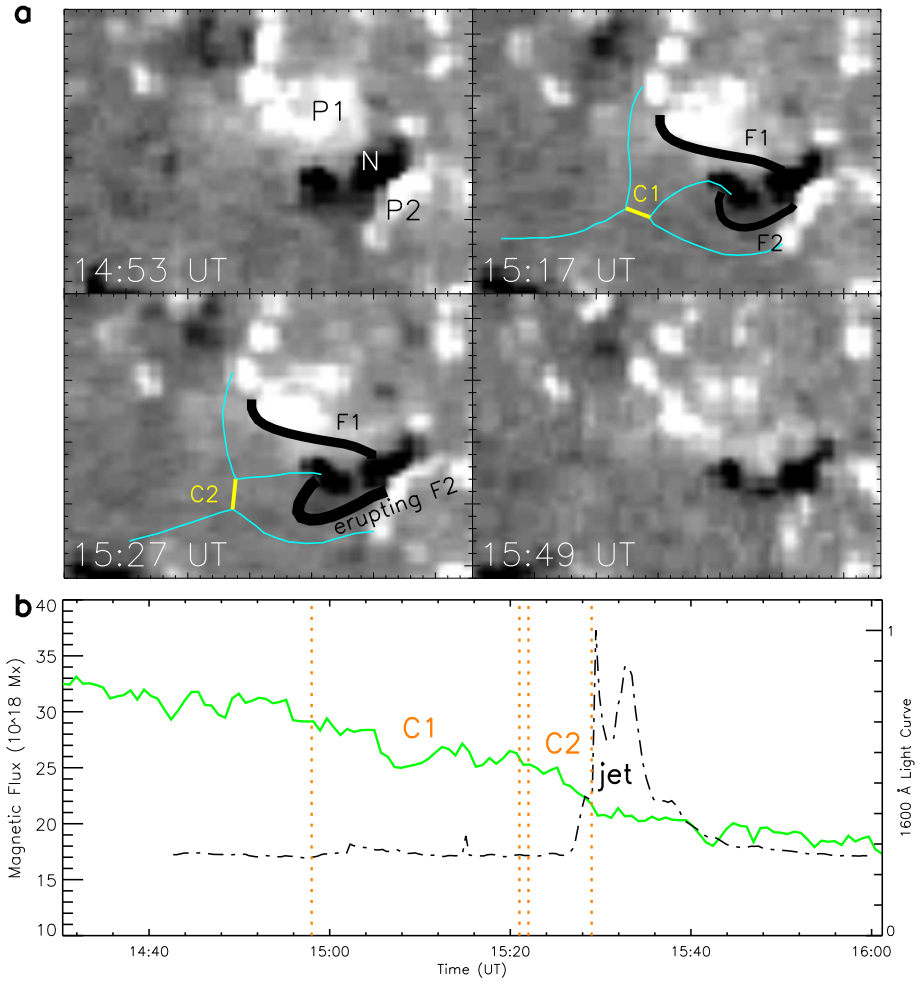}
\caption{{\bf a}: HMI magnetograms showing the photospheric magnetic field evolution in the jet base through before the formation of the breakout current sheet to the end of the jet.  The axises of F1/F2 (black curves), the current sheets C1/C2 (yellow lines) and their connectivity (cyan curves) to the magnetic polarities P1, P2 and N, are drew out along the corresponding {\it IRIS}/SJI 1400 \AA\ images, and are superimposed here on the magnetograms at 15:17 and 15:23 UT. {\bf b}: the temporal evolution of magnetic flux of N (green curve) showing a magnetic cancellation process, with the overlain AIA 1600 \AA\ light curve (dotted-dashed curve) in the jet base showing two peaks at 15:29:04 UT and 15:32:40 UT. The orange dotted lines have the same meaning as that in Figure 3, indicating the two duration of C1 and C2.}
\end{figure}

\end{document}